\documentclass[a4paper]{article}
\usepackage{ASVspoof}
\usepackage{epsfig,amssymb,amsmath}
\ninept

\usepackage{amsmath}
\usepackage{booktabs}
\usepackage{graphicx}
\usepackage{microtype}
\usepackage{xspace}
\usepackage{cite}
\usepackage{url}
\usepackage{tikz}
\usepackage{comment}
\usepackage{multirow}
\usepackage{xcolor}
\usepackage{pifont}
\usepackage{enumitem}
\usepackage[prependcaption,textsize=scriptsize]{todonotes}
\usepackage{tabularx}
\usepackage{cleveref}

\definecolor{indiagreen}{HTML}{138808}
\definecolor{papaya}{HTML}{EE892F}
\definecolor{mygreen}{HTML}{008000}
\definecolor{mypurple}{HTML}{9966CC}
\definecolor{mygray}{HTML}{696969}

\newcommand{\cmark}{\textcolor{black}{\ding{51}}}

\definecolor{mygreen}{HTML}{008000}
\definecolor{mypurple}{HTML}{9966CC}
\definecolor{myblue}{HTML}{5072A7}
\definecolor{myred}{HTML}{DC143C}
\definecolor{mypapaya}{HTML}{EE892F}

\title{WavLM model ensemble for audio deepfake detection}

\name{David Combei,$^1$ Adriana Stan,$^{1,2}$ Dan Oneață,$^2$ Horia Cucu~$^2$}

\address{
  $^1$Technical University of Cluj-Napoca, Romania\\
  $^2$\textsc{Politehnica} Bucharest, Romania \\
{\small \tt combei.ga.david@student.utcluj.ro, adriana.stan@com.utcluj.ro}\\{\small \tt dan.oneata@gmail.com, horia.cucu@upb.ro}
}

\begin{document}
\maketitle

\begin{abstract}
   
Audio deepfake detection has become a pivotal task over the last couple of years, as many recent speech synthesis and voice cloning systems generate highly realistic speech samples, thus enabling their use in malicious activities.
In this paper we address the issue of audio deepfake detection as it was set in the ASVspoof5 challenge.
First, we benchmark ten types of pretrained representations and show that the self-supervised representations stemming from the wav2vec2 and wavLM families perform best.
Of the two, wavLM is better when restricting the pretraining data to LibriSpeech, as required by the challenge rules.
To further improve performance, we finetune the wavLM model for the deepfake detection task.
We extend the ASVspoof5 dataset with samples from other deepfake detection datasets and apply data augmentation.
Our final challenge submission consists of a late fusion combination of four models and achieves an equal error rate of 6.56\% and 17.08\% on the two evaluation sets. 
\end{abstract}

\section{Introduction}
\label{sec:intro}
The capacity of generative deep learning has recently achieved remarkable results, and it has become close to impossible to perceptually distinguish between real (or \emph{bonafide}) and generated (or \emph{fake, spoofed}) data across multiple domains. Audio generation is no exception. High-quality text-to-speech (TTS) and voice cloning (VC) systems have become easily and readily available for all user categories. 
If the use of such technology is performed on one's behalf, for example to generate audio content for a video blog or online platform, its use supports the user tremendously and eases the process of generating vast amounts of online content. However, if these systems are used to impersonate or to alter an original audio or video resource, then the forensics of deepfake data should be supported by equally able detection systems.


\textbf{Audio deepfake detection.}
Modern deepfake detection increasingly relies on self-supervised representations \cite{wang22odyssey,donas22icassp,tak2022odyssey,yang2024icassp,guo2024icassp,wu2024icassp,saha24green,pascu24interspeech}.
Self-supervised learning (SSL) \cite{balestriero23ssl} is a powerful paradigm that aims to produce transferable representations.
Methods such as wav2vec~\cite{baevski20wav2vec2}, HuBERT~\cite{hsu21hubert} or wavLM~\cite{chen22wavlm}
achieve this desideratum by reconstructing masked parts of the input audio.
The resulting representations can be successfully employed by multiple downstream tasks (e.g. speech recognition, keyword spotting, speaker identification) with limited data \cite{yang21superb}.
This is also the case for our task of interest, audio deepfake detection,
where methods based on self-supervised representations provide a lighter \cite{saha24green} and more robust \cite{pascu24interspeech} alternative compared to the previous generation of approaches,
such as ResNet on linear frequency cepstral coefﬁcients (LFCC) \cite{chen21arxiv} or RawNet \cite{rawnet2}.

Most approaches use self-supervised models as feature extractors,
keeping them frozen during training  \cite{donas22icassp,yang2024icassp,wu2024icassp,pascu24interspeech,saha24green}.
Finetuning the self-supervised frontends has also been explored \cite{tak2022odyssey,wang22odyssey,guo2024icassp}, although to a lesser degree;
among these, Wang and Yamagishi \cite{wang22odyssey} suggest that finetuning the frontend reduces the reliance of the classifier on the type of backend.
The wav2vec family of models \cite{baevski20wav2vec2,conneau21xlsr,babu22xlsr2} remains the most popular option for deepfake detection \cite{donas22icassp,wang22odyssey,wang22damm,tak2022odyssey,xie23interspeech,saha24green,pascu24interspeech}.
WavLM and HuBERT representations have also been employed \cite{wang22odyssey,pascu24interspeech,guo2024icassp,yang2024icassp},
but their results seem to trail those of wav2vec.
Wav2vec comes in multiple variants and while most of the small versions have been employed \cite{saha24green,wang22odyssey},
the larger variants trained on more data perform best \cite{pascu24interspeech}.
Others have also combined these three types of features \cite{yang2024icassp} 
or used representations from earlier layers \cite{saha24green} or pooled information from multiple layers \cite{donas22icassp,guo2024icassp,wu2024icassp}.

Backends range from simple linear models \cite{pascu24interspeech,saha24green} to more complex pooling mechanisms \cite{wang22odyssey,donas22icassp,guo2024icassp,wu2024icassp}.
At one end of the spectrum, Pascu et al. \cite{pascu24interspeech} and Saha et al. \cite{saha24green} observe good results even for linear classifiers.
At the other end of the spectrum, Wang and Yamagishi \cite{wang22odyssey} suggest that more complex backends help more, with multi-fusion attention mechanism being a popular choice \cite{guo2024icassp,wu2024icassp}.


\textbf{Our work.} In this paper we address the topic of unimodal audio deepfake detection (or spoofing) in the context of the 2024 ASVspoof5 Challenge (ASV5)~\cite{asvpaper24}. 
The challenge was based on a very large crowdsourced dataset of spoofed audio samples generated with various TTS and VC systems. 
It contained two tracks: 1) deepfake detection; 2) automatic speaker verification.
For both tracks, closed and open conditions were also in place. 
The closed condition referred to using only the released data, thus restricting the use of other spoken samples or pretrained models.
In the open condition, there was a single limitation pertaining to the use of models or datasets which included samples from the LibriLight~\cite{kahn20librilight}, Multilingual LibriSpeech~\cite{pratap20mls} or MUSAN~\cite{musan2015} datasets.

Our challenge submissions and results address the \emph{open condition of the deepfake detection track}. As shown above, large SSL models have shown very good performance over the deepfake detection task. However, the limitation within the ASV5 challenge's open track discarded the majority of the top performing readily available pretrained models. 
As a result, we first explored additional SSL model families trained only on the LibriSpeech~\cite{panayotov2015librispeech} dataset, and benchmarked them as frontend feature extractors on a subset of the ASV5 data. We then selected the base variants of \texttt{wavLM} and \texttt{wav2vec2} and finetuned their parameters for the deepfake detection task. Our final challenge submission aggregated the predictions of several pretrained and finetuned models and obtained a 17.08\% EER.


\section{Benchmarking pretrained model representations}
\label{sec:bench}

We investigate how well pretrained audio representations transfer to the task of audio deepfake detection.
We consider three classes of models.
The first class are self-supervised models, which are pretrained on unlabelled data:

\begin{itemize}[leftmargin=*]

    \item DeCoAR2~\cite{ling2020decoar2} is a Deep Contextualized Acoustic Representation model using vector quantisation. The model was trained on 960h of LibriSpeech.
    
    \item HuBERT~\cite{hsu21hubert} is a Hidden-Unit BERT approach for self-supervised speech representation learning.
    It was trained on LibriSpeech and finetuned on different subsets.
    
    \item Distill-HuBERT~\cite{chang2022distilhubert} is a version of HuBERT pretrained SSL model, that reduces its size by 75\%. This model was trained for several downstream tasks using SUPERB dataset.
    
    \item wavLM~\cite{chen22wavlm} learns masked speech prediction and it denoises the data during training to enhance the performance. The base version is trained on 960h of LibriSpeech data.
    
    \item wav2vec2.0~\cite{baevski20wav2vec2} is a well known framework for self-supervised learning of speech representations, it masks the speech input in the latent space and solves a contrastive task defined over a quantisation of the latent representations which are jointly learned.
    The base version was trained on LibriSpeech data.

    \item BEATs~\cite{chen2023beats} is an iterative audio pretraining framework to learn Bidirectional Encoder representation from Audio Transformers, where an acoustic tokenizer and an audio SSL model are iteratively optimised. This model was trained on Audioset-2M~\cite{audioset} which also includes non-speech audio.   
    
\end{itemize}

The second class of models are speaker embedding networks.
We selected this category of models, since they were shown to capture other information as well \cite{math10213927}.
We selected the ECAPA-TDNN~\cite{desplanques2020ecapatdnn} and TitaNet~\cite{koluguri2022titanet} models.
ECAPA-TDNN is a time delay neural network that applies statistics pooling to project variable-length utterances into fixed-length speaker embeddings;
this model was trained on the VoxCeleb dataset.
For TitaNet we use its \texttt{large} variant, which was trained for speaker verification and diarisation on tens of thousands of hours of audio data from VoxCeleb 1, VoxCeleb 2, Fisher, Switchboard, LibriSpeech and SRE dataset.

Finally, the third class of models is that of learnable frontends.
LEAF~\cite{zeghidour2021leaf} was created to replace mel-filterbanks for audio classification of speech, music, audio events and animal sounds. 
HEAR's YAMNet~\cite{turian2021hear} is also trained for audio classification using a knowledge distillation approach with transformers and CNNs.
Both models were trained on the Audioset data.

As topline, we also include results for two models trained on LibriLight \cite{kahn20librilight} or Multilingual LibriSpeech \cite{panayotov2015librispeech}, which were consequently not allowed in the challenge:
\texttt{wavLM-large} and \texttt{wav2vec2-xls-r-2b}. 
The large variant of WavLM is a three times larger model than the base one, trained on 94k hours of speech (60k LibriLight, 10k Giga-Speech, 24k VoxPopuli).
The XLS-R 2B variant of wav2vec~\cite{babu22xlsr2} is a large-scale model for cross-lingual speech representation learning based on wav2vec 2.0.
This model was trained on nearly half a million hours of publicly available speech audio in 128 languages.

\begin{table}[t!]
    \newcommand{\ii}[1]{{\footnotesize \color{gray} #1}}
    \centering
    \caption{
    Performance of self-supervised representations in terms of equal error rate (EER) on a subset of 27k samples from the ASV5 development set.
    Last two models are pretrained on either LibriLight or Multilingual LibriSpeech, and hence they do not adhere to the challenge rules.
    Lower values are better.
    The models are listed in decreasing order of their performance and have associated information regarding their parameter count and extracted feature's dimension.
    } \vspace{0.1cm}
    \begin{tabular}{rlrrr}
    \toprule
     & \textbf{Model} & \# $param$ & $feat$ & EER $\downarrow$ \\
     &                &             &  $dim$     & \multicolumn{1}{c}{[\%]} \\
     \midrule
        \ii{1}  & LEAF~\cite{zeghidour2021leaf}               & 4M & 40    &    50.14 \\
        \ii{2}  & Distill-HuBERT~\cite{chang2022distilhubert} & 23M & 768  &    32.37 \\
        \ii{3}  & ECAPA-TDNN \cite{desplanques2020ecapatdnn}  & 6M & 192  &    28.23 \\
        \ii{4}  & HEAR's YAMNet~\cite{turian2021hear}         & 4M & 184  &    23.69 \\
        \ii{5}  & TitaNet-large~\cite{koluguri2022titanet}    & 23M & 192  &    20.84 \\
        \ii{6}  & BEATs~\cite{chen2023beats}                  & 90M & 768  &    19.23 \\
        \ii{7}  & DeCoAR2~\cite{ling2020decoar2}              & 85M & 768  &    18.74 \\
        \ii{8}  & HuBERT~\cite{hsu21hubert}                 & 95M & 768  &    16.47 \\
        \ii{9}  & wav2vec2-base~\cite{baevski20wav2vec2}      & 94M  & 768  &    13.33 \\
        \ii{10} & wavLM-base~\cite{chen22wavlm}               & 94M  & 768  & \bf 9.93 \\
        \midrule
        \multicolumn{5}{l} {\color{gray}{\scriptsize Models pretrained on LibriLight or Multilingual Librispeech }} \\
        \ii{11} & wavLM-large~\cite{chen22wavlm}              & 300M & 1024 & \color{gray} 6.67 \\
        \ii{12} & wav2vec2-xls-r-2b~\cite{babu22xlsr2}        & 2B   & 1920 & \color{gray} 0.96 \\
    \bottomrule
    \end{tabular}

\label{tbl:ssl-models}
\end{table}

\begin{figure}[t!]
  \centering
  \includegraphics[width=\columnwidth,trim={120pt 30pt 120pt 30pt},clip]{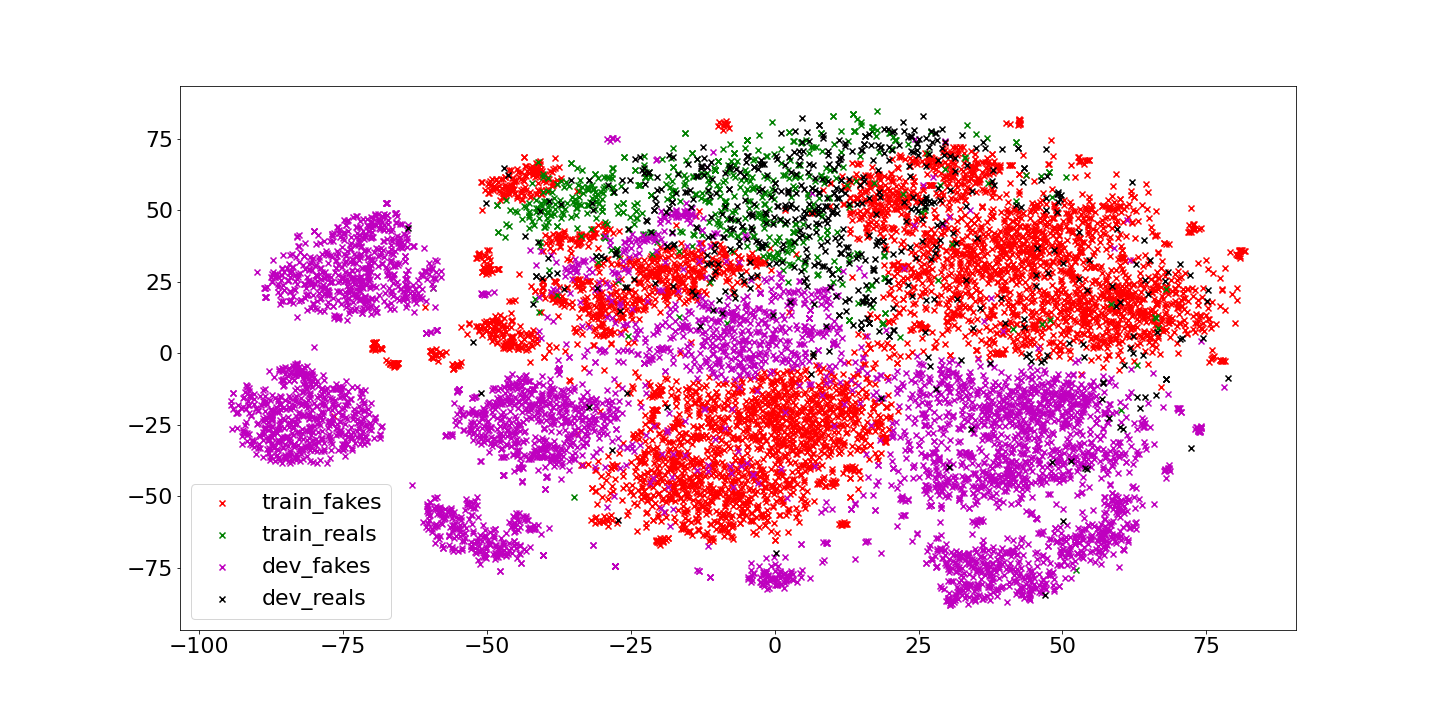} \\(a) \scriptsize{wavLM-base}
  \includegraphics[width=\columnwidth,trim={120pt 30pt 120pt 30pt},clip]{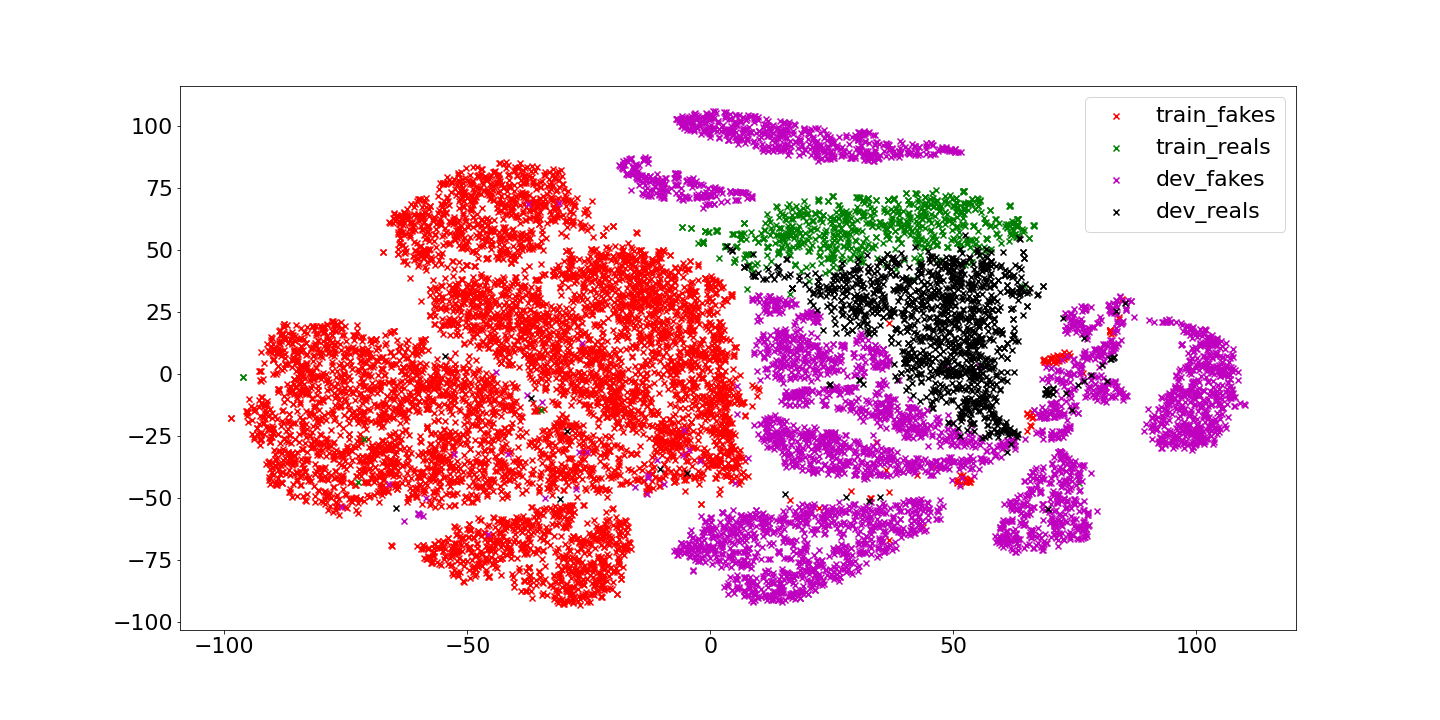}\\(b) \scriptsize{wav2vec2-xls-r-2b}
  \caption{%
    t-SNE plots of the pretrained (a) \texttt{wavlm-base}, (b) \texttt{wav2vec2-xls-r-2b} representations on a random subset of the ASV5 training and development data.
  }
  \label{fig:tsne-wavlm}
\end{figure}

To get a grasp of the models' inherent capabilities, we plot t-SNE projections of the \texttt{wavLM-base} and \texttt{wav2vec2-xls-r-2b} representations for a subset of the ASV5 data (see Figure~\ref{fig:tsne-wavlm}). It can be noticed that the \texttt{wav2vec2-xls-r-2b} features exhibit a clear separation between the four subsets of data:
spoofed samples from train (red), bonafide samples from train (green), spoofed samples from dev (purple), bonafide samples from dev (black).
This indicates that this self-supervised representation is powerful enough to discriminate spoofed from bonafide samples even when simple backend classifiers are employed.
Moreover, we observe distinct clusters even inside the spoofed data,
presumably corresponding to each of the eight attacks in training and development sets. 
For the \texttt{wavLM-base} representation, some similar clusters emerge, but their separation hyperplanes are not as clearly defined. It is worth mentioning that the t-SNE projection is non-linear, and that the \texttt{wavLM} features may still exhibit linearly separable clusters in the $N$-dimensional space.

For audio deepfake detection benchmarking we selected two random subsets of 27k samples from each of the training and development sets of the ASV5 Challenge data, respectively. 
Both subsets have the same data statistics as the original sets, i.e., 8 spoof files to 1 bonafide file.
We use the SSL models as feature extractors (without finetuning)
and evaluate their performance for deepfake detection by linear probing.
Specifically, we train a logistic regression model over the average pooled representations to predict the `bonafide' or `spoof' label.
The logistic regression uses a regularisation term%
\footnote{As defined by the \texttt{C} parameter in the scikit-learn documentation}
of $10^3$.

Table~\ref{tbl:ssl-models} shows these results.
We observe a wide range of performances, from 50\% EER (random chance) to less than 10\% EER, which is obtained by the base version of the \texttt{wavLM} model.
Models pretrained on LibriLight (\texttt{wavLM-large}) or Multilingual LibriSpeech (\texttt{wav2vec2-xls-r-2b}) improve the performance even further to 0.96\% EER for the later.
But since these models were not allowed in the competition, we focus on the \texttt{wavLM-base} model and proceed to finetune its parameters on the audio deepfake task.

\section{Model finetuning}
\label{sec:finetuning}

In the previous section we found that the pretrained \texttt{wavLM-base} and \texttt{wav2vec2-base} models exhibited the best deepfake detection performance within the ASV5 challenge's limitations. To further increase the models' discriminative power, we perform a light finetuning of their weights. For this we also use external audio data and signal-based data augmentation. The finetuned models were again used as feature extractors, and a logistic regressor provided the final classification labels.

\subsection{Additional audio data and data augmentation}
\label{subsect:data}

The initial phase of the ASV5 challenge released a set of $182,357$ training samples ($18,797$ bonafide | $163,560$ spoofed); $140,950$ development samples ($31,334$ bonafide | $109,616$ spoofed), and a small progress phase evaluation set of $40,765$ samples (no labels were given for this subset). The final evaluation was performed over $680,774$ samples. From our initial data analysis, we assumed that one of the challenge's objectives was to explore the models' generalisation abilities. Therefore, we also included in the training data a random selection of samples from other audio deepfake datasets, as follows:

\begin{itemize}[leftmargin=*]
    \item{\textbf{ASVspoof 2019} (ASV19) \cite{WANG2020101114} -- from all subsets;}
    \item{\textbf{ASVspoof 2021} (ASV21) \cite{asvspoof21} -- from the evaluation subset;}
    \item{\textbf{Fake or Real} (FoR) \cite{for} -- from the `norm' subset; }
    \item{\textbf{In the Wild} (ITW) \cite{muller2022interspeech} -- from all.}
\end{itemize}

We also explored signal-based data augmentation. Specifically, we added white noise\footnote{ With a signal to noise ratio of 25dB} and reverberation\footnote{ As described in the \texttt{torchaudio} tutorial: \url{https://pytorch.org/audio/stable/tutorials/audio_data_augmentation_tutorial.html}} to the bonafide files in ASV5; we applied a single random augmentation to an audio sample; half of the bonafide samples were augmented. We did not alter the spoofed samples. 

Based on the above steps, we obtained three dataset variants which were used for finetuning:
\begin{itemize}[leftmargin=*]
    \item \texttt{medium-27k} consists of a subset of 27k samples from ASV5 training set having the same distribution as the original set (8 spoofed files to 1 bonafide file, equal number of files from each spoofing attack)--same as in Section~\ref{sec:bench};
    \item \texttt{augm-31k} consists of 31k samples: 13k samples from ASV5 with augmentation for half of the bonafides, 6.1k from ASV19, 8.6k from ASV21, 1.6k from ITW, 1.8k from FoR;
    \item \texttt{augm-114k} consists of 114k samples: 102k samples from ASV5 
    with augmentation for half of the bonafides, 2.9k from ASV19, 6.8k from ASV21, from 1.6k FoR, 1.6k from ITW.
\end{itemize}

The \texttt{augm-31k} was selected to have a similar number of samples as the \texttt{medium-27k} set. While the \texttt{augm-114k} subset was chosen to explore if more finetuning data increases the model's ability to generate more discriminative features for the deepfake detection task.\footnote{Given our limited computational resources, we did not attempt to finetune the models using the complete datasets.}

In the numeric evaluation we used the same 27k subset of the development set as in Section~\ref{sec:bench} and the small progress phase evaluation set provided by the organisers.

\subsection{Implementation details}

The models were finetuned to minimise the binary cross-entropy loss over the deepfake detection task.
The Adam optimiser was used with a learning rate of $3\cdot 10^{-5}$
and a linear scheduler with a warm up ratio of 0.1 over .
We finetuned the models over 5 epochs using a batch size of either 8 (for \texttt{medium-27k} and \texttt{augm-114k}) or 16 (for \texttt{augm-31k}).
Training was performed on a single Tesla V100 16GB GPU and took around 20 hours for the \texttt{augm-114k} split.

The finetuning process yielded 3 variants of the \texttt{wavLM-base} model corresponding to the three datasets described in the previous section, and one variant of the \texttt{wav2vec2-base} model finetuned only with the \texttt{medium-27k} subset.

\subsection{Results}

The results are shown in \Cref{tbl:finetuning} and indicate that finetuning improves over the pretrained representations.
Among the three \texttt{wavLM-base} finetuning variants, the ones using augmented data perform best:
the \texttt{agum-31k} set gives the lowest error on the development set (0.61\% EER),
while the \texttt{augm-114k} set gives the lowest error on the progress phase evaluation set (7.26\% EER).

For the \texttt{wav2vec2-base} representation we observe similar improvements over the pretrained variant.
However, the performance remains worse in the absolute than that of the \texttt{wavLM-base} model:
11.91\% EER for the finetuned wav2vec model (row 6) versus 7.26\% EER for the best finetuned wavLM model (row 4).

\begin{table}[th!]
    \newcommand{\ii}[1]{{\small \color{gray} #1}}
    \centering
    \caption{
        EER [\%] performance of finetuned \texttt{wavLM} and \texttt{wav2vec2} models on the ASV5 development and progress phase evaluation sets.
    }
    \vspace{0.1cm}
    \label{tbl:finetuning}
    \begin{tabular}{rlcrr}
    \toprule
                        &            &              &  \multicolumn{2}{c}{EER[\%] $\downarrow$} \\
                        & Model type & Training set & Dev & Prog \\
      \midrule
      \multicolumn{5}{l}{\color{darkgray}{\scriptsize wavLM variants: \texttt{wavlm-base}}} \\
      \ii{1}   & Pretrained   & --                  & 9.93         & 15.82    \\
      \ii{2}   & Finetuned    & \texttt{medium-27k} & 4.16         & --       \\
      \ii{3}   & Finetuned    & \texttt{augm-31k}   & \bf 0.61     & 9.02     \\
      \ii{4}   & Finetuned    & \texttt{augm-114k}  & 2.97         & \bf 7.26 \\
      \midrule
      \multicolumn{5}{l}{\color{darkgray}{\scriptsize wav2vec2 variants: \texttt{wav2vec2-base}}}  \\
      \ii{5}   & Pretrained  & --                    & 13.33          & --    \\
      \ii{6}   & Finetuned   & \texttt{medium-27k}   & 5.85           & 11.91 \\
    \bottomrule
    \end{tabular}
    \vspace{-0.4cm}
\end{table}

\begin{table}
\centering
\newcommand{\ii}[1]{{\small \color{gray} #1}}
\caption{%
    EER [\%] performance on development, progress phase evaluation and final evaluation sets for
    the late fusion combinations of four \texttt{wavLM-base} models (see Table~\ref{tbl:finetuning}).
    The models vary by type (pretrained -- PT or finetuned -- FT) and data used for finetuning (\texttt{medium-27k}, \texttt{augm-31k} or \texttt{augm-114k}). Lower values are better. Best results are marked in boldface.
}
\vspace{0.1cm}
\label{tbl:late-fusion}
\begin{tabular}{rcccc|rrr}
    \toprule
           &            &              &          & &\multicolumn{3}{c}{EER[\%] $\downarrow$} \\
    Type:  & PT     & FT           & FT           & FT            & & &  \\
    Data:  & --     & \texttt{27k} & \texttt{31k} & \texttt{114k} & Dev & Prog & Eval \\
    \midrule                                        
    \ii{1} &        &              &              & \cmark        &  2.97          & 7.26          & --             \\ 
    \ii{2} & \cmark &              &              & \cmark        &  1.16          & --            & --             \\ 
    \ii{3} &        & \cmark       &              & \cmark        &  1.17          & --            & --             \\ 
    \ii{4} &        &              & \cmark       & \cmark        &  \textbf{0.56} & --            & --             \\ 
    \ii{5} &        & \cmark       & \cmark       & \cmark        &  0.60          & --            & --             \\ 
    \ii{6} & \cmark & \cmark       & \cmark       & \cmark        &  0.72          & \textbf{6.56} & \textbf{17.08} \\ 
    \bottomrule
\end{tabular}
\vspace{-0.4cm}
\end{table}

\begin{figure}[t!]
  \centering
  \includegraphics[width=\columnwidth,trim={120pt 30pt 120pt 30pt},clip]{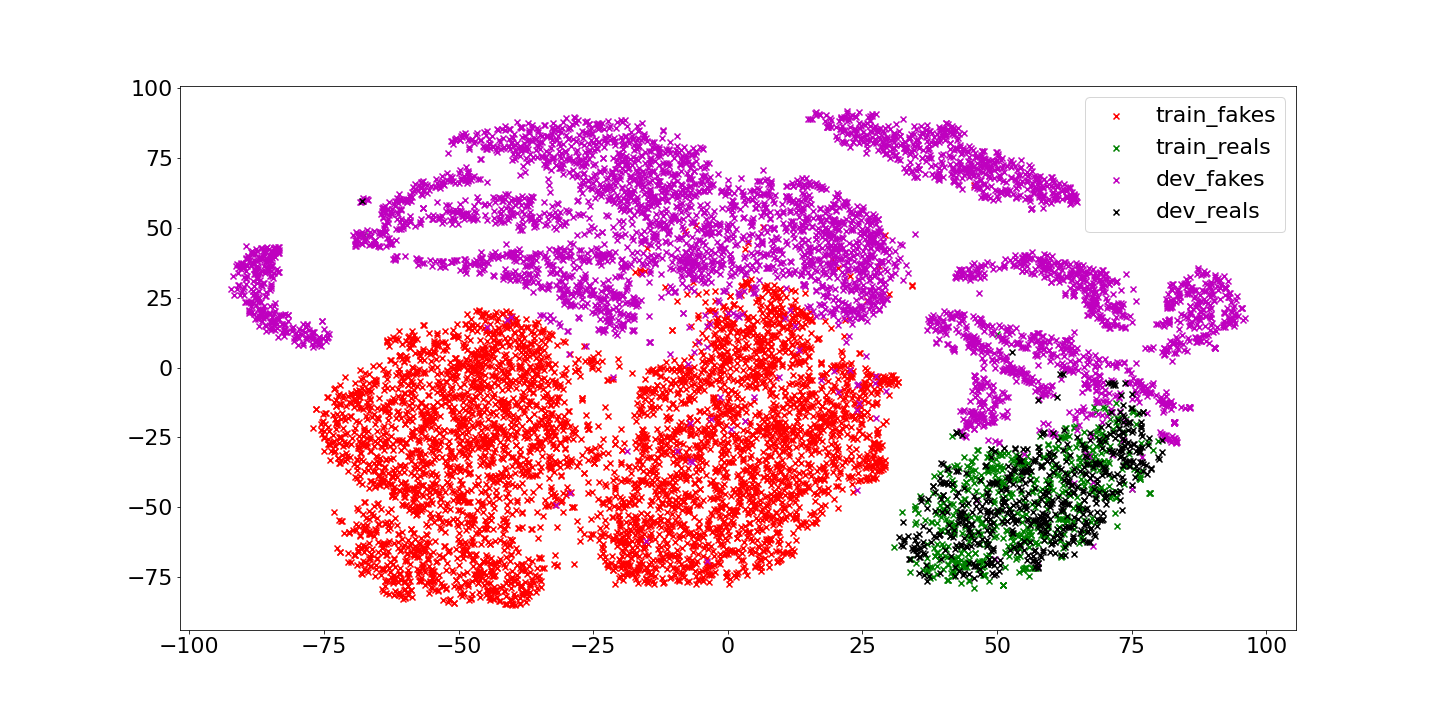} \\(a) \scriptsize{wavLM-finetuned on medium-27k}
  \includegraphics[width=\columnwidth,trim={120pt 30pt 120pt 30pt},clip]{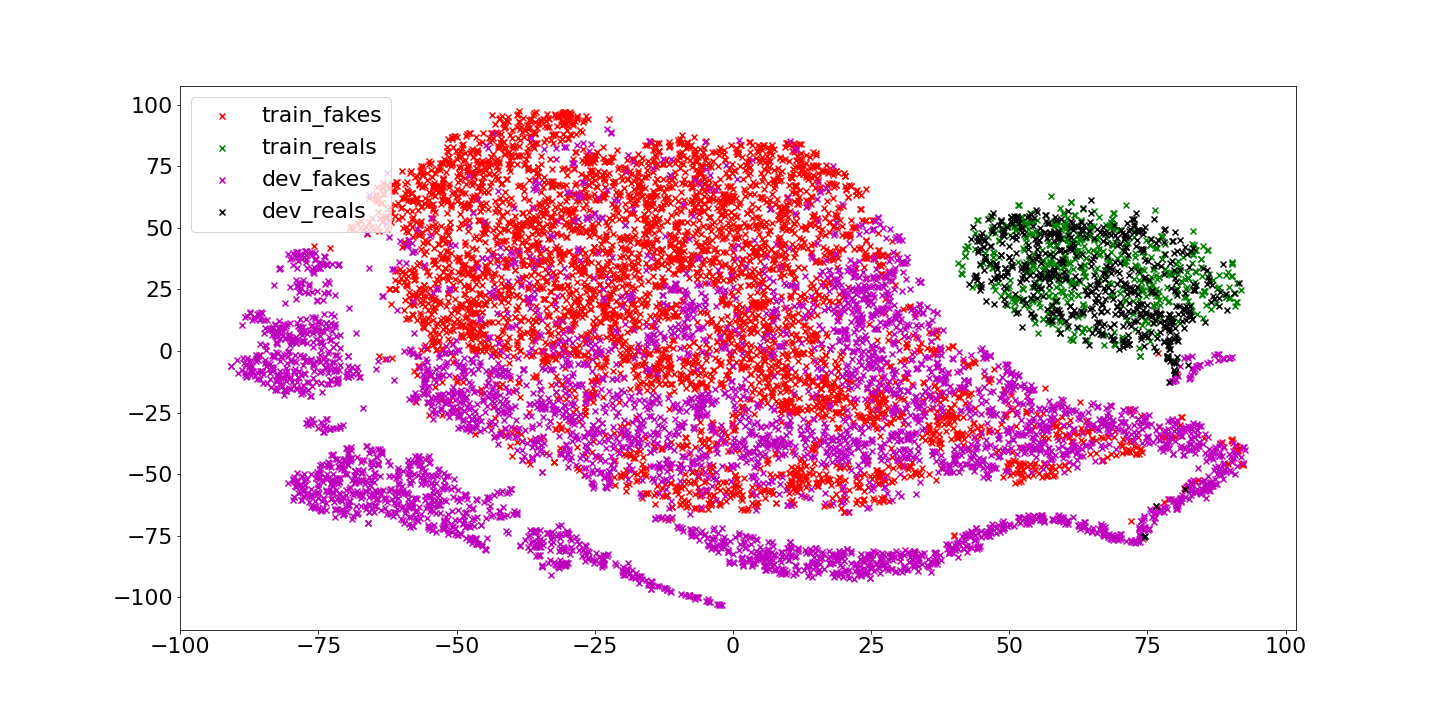} \\(b) \scriptsize{wavLM-finetuned on augm-31k}
  \includegraphics[width=\columnwidth,trim={120pt 30pt 120pt 30pt},clip]{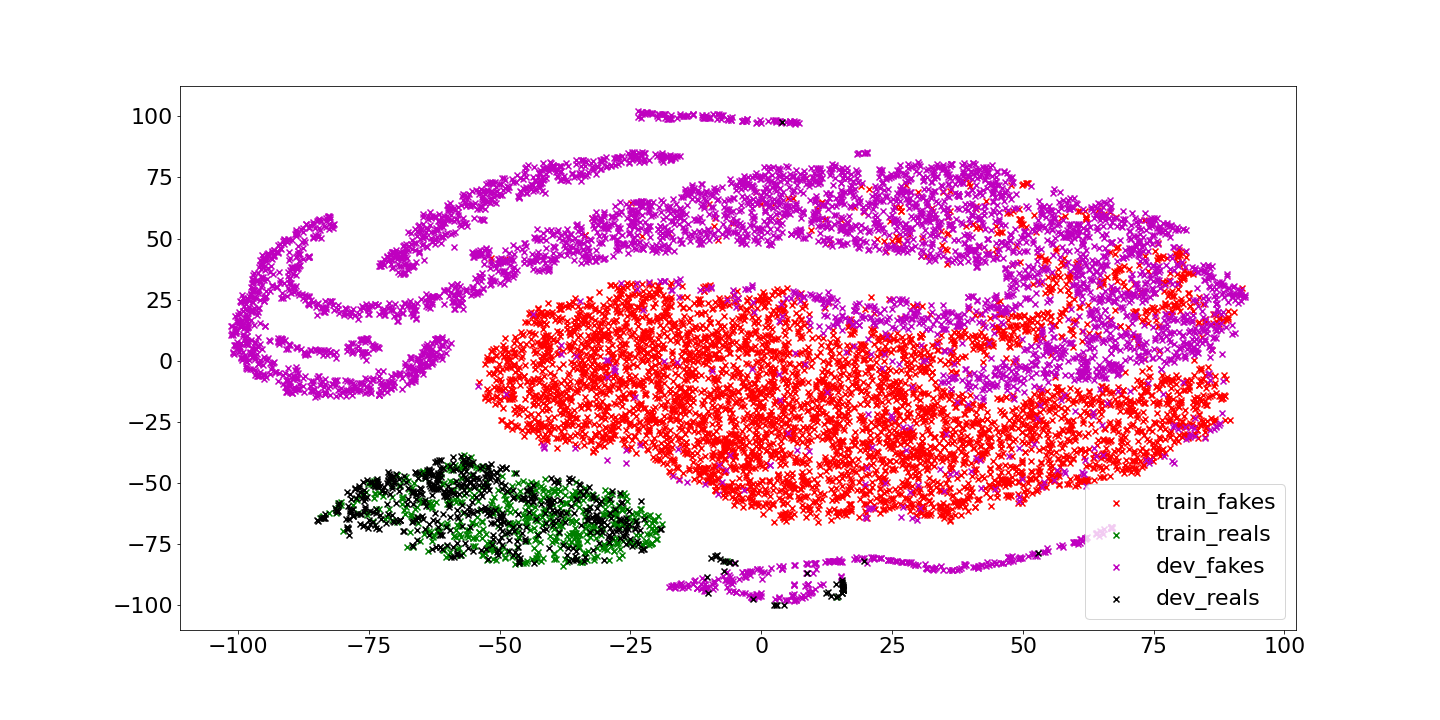}\\(c) \scriptsize{wavLM-finetuned on augm-114k}
  \caption{%
    t-SNE plots of the representations obtained from the finetuned wavLM models using the (a) \texttt{medium-27k}, (b) \texttt{augm-31k}, (c) \texttt{augm-114k} datasets.
  }
  \label{fig:tsne2}
\end{figure}

We again examined the t-SNE plots of the features extracted from the finetuned models and show them in Figure~\ref{fig:tsne2}. It can be noticed that, as opposed to Figure~\ref{fig:tsne-wavlm}a, the separation between the real and fake samples is more clearly defined. However, the different attacks still do not seem as clustered as for the \texttt{wav2vec2-xls-r-2b} features (see Figure~\ref{fig:tsne-wavlm}b).

\section{Model ensemble}

For  the final submission to the ASV5 Challenge, we combined the predictions of multiple models using late fusion.
We learn a set of weights using separate logistic regression over the probabilities output by the models presented in \Cref{sec:finetuning}.
The late fusion weights are unconstrained, so they can be either positive or negative.
To learn the weights, we used the \texttt{medium-27k} dataset train split.

The results are shown in \Cref{tbl:late-fusion}.
We start from the best performing model found in the previous section (listed in row 1 of the table), and
first combine it with one (rows 2--4), then two (row 5) and finally all of the others models (row 6).
On the development set, we observe improvements by using model combinations, with the best combination of two consisting of the models finetuned on the largest datasets: \texttt{augm-31k} and \texttt{augm-114k} (row 4).
Our submission to the challenge corresponds to the combination of all four variants (row 6) and yields an EER of 6.56\% on the progress phase evaluation set, and 17.08\% on the final evaluation set.\footnote{Not all results over the progress phase evaluation set could be presented here, as the phase was closed by the time of the current submission.}

\section{Conclusions}
\label{sec:conc}


This paper presented our submission to the ASVspoof 2024 deepfake detection challenge.
First, we benchmarked a set of ten pretrained representations belonging to three classes of models (self-supervised, speaker embedding, learnable frontends) and have shown that the self-supervised models outperform the others, with wavLM achieving the best performance.
Second, we have shown that we can further improve the performance of the pretrained representations by finetuning them for the task of deepfake detection.
When finetuning we found data augmentation to be an important component.
Our final submission combined four different models (pretrained and finetuned) in an ensemble and delivered the best performance.

\section{Acknowledgements}
This work was funded by EU Horizon projects AI4TRUST (No. 101070190) and by CNCS/CCCDI UEFISCDI (No. PN-IV-P8-8.1-PRE-HE-ORG-2023-0078).

\bibliographystyle{IEEEtran}
\bibliography{mybib}

\end{document}